# Pulsed laser deposition of ZnO thin films decorated with Au and Pd nanoparticles with enhanced acetone sensing performance

M. Alexiadou, M. Kandyla·, G. Mousdis, and M. Kompitsas

*National Hellenic Research Foundation, Theoretical and Physical Chemistry Institute, 48 Vasileos Constantinou Avenue, 11635 Athens, Greece*

**Abstract**

We fabricate and compare nanocomposite thin-film ZnO chemoresistive acetone sensors with gold or palladium nanoparticles on the surface, at low operating temperatures. The sensors are fabricated by pulsed laser deposition and operate in the temperature range 159 – 200$^o$C. The ZnO films are polycrystalline, crystallizing mainly at the (002) and (101) orientations of the hexagonal phase. The nanocomposite ZnO:Au and ZnO:Pd sensors have a lower detection limit and show a response enhancement factor between 2 – 7, compared with pure ZnO sensors. The ZnO:Pd sensor performs better than the ZnO:Au sensor. The ZnO:Pd sensor sensitivity increases with the amount of palladium on the surface, while it remains roughly unchanged with the ZnO thickness. The lowest acetone concentration we detect is 26 ppm for the operating temperature of 200$^o$C.

· Corresponding author:
M. Kandyla
National Hellenic Research Foundation
Theoretical and Physical Chemistry Institute
48 Vasileos Constantinou Avenue
11635 Athens, Greece
Tel.: +30 210 7273826, Fax: +30 210 7273794
E-mail: kandyla@eie.gr

# 1. Introduction

Acetone is a flammable volatile organic compound, which is widely used as an industrial solvent. However, it can affect the central nervous system as well as the liver and kidney. Additionally, acetone is contained in the human breath and may act as a biomarker for disease diagnostics, such as type-1 diabetes [1] and lung cancer [2]. Also, it may enable accurate monitoring of blood glucose levels for non-invasive and efficient diabetic patient self-management. Therefore, sensitive and reliable acetone sensors are essential for medical, environmental, and safety applications.

Sensing of acetone has been achieved by metal-oxide chemoresistive sensors, such as $WO_3$ [3,4], $Fe_2O_3$ [5], $In_2O_3$ [6], $SnO_2$ [7], ZnO [8], and their heterostructures [9], among others. Metal-oxide gas sensors can be easily developed by a variety of simple fabrication methods and offer high sensitivity and reduced cost. ZnO is a promising metal oxide for gas sensing applications, due to low cost, chemical and thermal stability, and non-toxicity. However, high operating temperatures are usually required in order to obtain highly sensitive ZnO sensors. In an effort to improve the sensing characteristics of the material, porous [10,11] or doped [12,13,14] ZnO, as well as ZnO microflowers [15], nanorods [16,17], nanoparticles [18,19], and quantum dots [20] have been employed recently as acetone sensors, among others.

Complementary to doping [12,21], the deposition of metallic nanoparticles on metal-oxide sensing surfaces has been shown to improve the sensing characteristics of the devices, lowering the detection limit, increasing the sensitivity, decreasing the response time, and lowering the operating temperature [22,23,24,25]. ZnO microflowers, decorated with gold nanoparticles prepared by chemical synthesis, were able to detect 5 – 10 ppm of acetone with high sensitivity and short response time, for operating temperatures of 270 – 280°C [22,23]. ZnO nanostructures, decorated with gold and palladium nanoparticles by sputtering, were employed for the detection of 500 ppm and 1000 ppm of acetone at 300°C [26].

In this work, we compare the performance of pure thin-film ZnO acetone sensors with nanocomposite ZnO:Au sensors (ZnO thin films with a surface partially covered by gold nanoparticles) and ZnO:Pd sensors (ZnO thin films with a surface partially covered by palladium nanoparticles) for low operating temperatures. We fabricate all sensors by pulsed laser deposition (PLD). PLD is a versatile technique with a high instantaneous deposition rate, which allows for the formation of thin films with controllable thickness as well as optical and electrical properties [27,28]. The sensors operate in the temperature range 159 – 200°C. The nanocomposite ZnO:Au and ZnO:Pd sensors have a lower detection limit than the pure ZnO sensor, especially for lower operating temperatures. The nanocomposite sensors are also more sensitive than the pure ZnO sensor, showing a response enhancement factor between 2 – 7 for similar operating temperatures and acetone concentrations. The ZnO:Pd sensor performs better than the ZnO:Au sensor and the sensing mechanism is discussed. A study on the ZnO film thickness and the amount of palladium on the ZnO surface reveals the ZnO:Pd sensor sensitivity increases with the amount of palladium on the surface, while it remains roughly unchanged with the ZnO thickness. The lowest acetone concentration we detect is 26 ppm for the operating temperature of 200°C.

## 2. Materials and Methods

Thin ZnO films were prepared by reactive pulsed laser deposition [29-31]. A zinc target was laser irradiated in a vacuum chamber under an oxygen flow with a dynamic pressure of 20 Pa. Prior to deposition, the chamber was evacuated to a base pressure of $10^{-5}$ mbar. The laser source was a Q-switched Nd:YAG system (pulse duration 10 ns, repetition rate 10 Hz), operating at 355 nm. The incident laser fluence on the zinc target was 6.5 – 7 J/cm$^2$. A glass substrate, on which the ZnO films were deposited, was placed 45 – 50 mm away from the zinc target and heated to 300$^o$C during deposition. Two irradiation times, of 120 min and 180 min were employed, resulting in samples of different thickness.

Metallic nanoparticles (gold and palladium) were deposited on the surface of the films on a second PLD process. This time, the ZnO films on glass were employed as substrates, heated to 120$^o$C, and gold or palladium targets were irradiated by the Nd:YAG laser beam for 1 or 2 min in vacuum, keeping the other deposition parameters the same as described above. This way, metallic nanoparticles were dispersed on the surface of the films, without completely covering them. The quantity of the nanoparticles increases with deposition time. Structural characterization of the samples was performed by a Theta D5000 X-ray diffractometer (XRD) with Cu Kα radiation. The composition of the films was determined by Energy Dispersive Analysis of X-rays (EDS).

The ZnO:Au and ZnO:Pd films were employed as gas sensors and their response to acetone was investigated in a home-built sensing setup, under static gas pressure conditions. Commercial acetone (ACS reagent) may contain up to 0.5% water, which can compromise the sensing measurements. In order to obtain water-free acetone, 10 gr of $Na_2SO_4$ were added in 200 ml of acetone and the mixture was kept in a sealed vial for 48 hours. The acetone was decanted to a vial and distilled under $N_2$. The first 20 ml were rejected and the following 50 ml were collected to a pre-dried vial (at 120° for 4 h) and sealed.

The ZnO:Au and ZnO:Pd films were annealed at 200°C for five hours before acetone sensing measurements. The samples were placed in a vacuum chamber in which a mixture of acetone vapor, diluted in nitrogen in a pre-mixing chamber, and dry air was introduced. The acetone concentration in air was calculated from the acetone/nitrogen partial pressures in the chamber, measured by an MKS Baratron gauge, taking into account the acetone dilution factor in nitrogen. The electric current through the samples was recorded in real time by a Keithley 485 picoammeter at a bias voltage of 1 V. During measurements, the films were heated between 159 and 200°C and their temperature was continuously recorded using a thermocouple. The sensor response, $S$, was calculated as

$$S = (R_o - R_g)/R_o \qquad (1),$$

where $R_g$ is the electric resistance of the sample in the presence of acetone and $R_o$ is the resistance of the sample in air.

## 3. Results and Discussion

Figure 1 shows an X-ray diffractogram of a ZnO film, deposited by a 2-hour long PLD process. The film is polycrystalline, with two peaks at $2\theta = 34.51°$ and $36.35°$, which are attributed to crystallization of the film at the (002) and (101) orientations of the hexagonal wurtzite phase, respectively, according to the JCPDS Card No. 36-1451. The average size of the ZnO grains, $D$, is estimated as ~54 nm from the Scherrer equation:

$$D = \frac{0.9 \times \lambda}{FWHM \times \cos\vartheta} \qquad (2)$$

where $\lambda = 1.5418$ Angstrom (CuK$\alpha$), $\theta$ is the central XRD peak angle, and FWHM is the full width at half maximum of the XRD peak. The X-ray diffractograms for the nanocomposite ZnO:Au and ZnO:Pd films do not show any peaks associated with gold or palladium, probably due to the small quantity of these materials on the ZnO surface. Indeed, EDS measurements in the Supplementary Material show the composition of the films with the highest Au or Pd quantities, which are (a) a ZnO film, formed by a 2-hour long PLD process, with gold nanoparticles on the surface, resulting from a 1-min irradiation of the gold target and (b) a ZnO film, formed by a 3-hour long PLD process, with palladium nanoparticles on the surface, resulting from a 2-min irradiation of the palladium target. Further information about the morphology of the films can be found in Ref. 31, which describes the properties of pulsed laser-deposited ZnO:Au thin films with similar experimental conditions.

A typical set of real-time measurements is shown in Fig. 2, which presents the transient response of a ZnO:Pd sensor to various concentrations of acetone in air, for an operating temperature of 185$^{o}$C. The sensor consists of a ZnO film, formed by a 3-hour long PLD process, with palladium nanoparticles on the surface, resulting from a 2-min irradiation of a palladium target. The sensor response increases in the presence of acetone and returns to equilibrium values upon flushing the chamber with dry air. According to the response definition in Eq. 1, this indicates that the electric resistance of the sensor decreases in the presence of acetone. Figure 2 shows that the sensor clearly responds to acetone concentrations as low as 31 ppm at this moderate operating temperature, with a very high signal-to-noise ratio. We also observe that flushing between detection cycles does not always result in the same equilibrium values and the equilibrium baseline seems to increase with the sensor operation. This is a common effect in metal-oxide gas sensors, which is due to the fact that, after each detection cycle, remaining acetone molecules which are not completely removed upon flushing render some sites of the sensor surface inactive.

In order to compare the performance of sensors with different surface materials, we examine the sensor calibration curves, shown in Fig. 3. This Figure presents the sensor response as a function of the acetone concentration for three different sensors: (a) a pure ZnO film, formed by a 2-hour long PLD process, (b) a similar ZnO film with gold nanoparticles on the surface, resulting from a 1-min irradiation of the gold target, and (c) a similar ZnO film with palladium nanoparticles on the surface, resulting from a 1-min irradiation of the palladium target. For all sensors we compare the response for two different operating temperatures, 175$^o$C and 200$^o$C. We observe the sensors with metallic nanoparticles on the surface (Fig. 3b and c) allow for the detection of smaller acetone concentrations, compared with the pure ZnO sensor (Fig. 3a), especially for the lower operating temperature of 175$^o$C. For similar operating temperatures and acetone concentrations the ZnO:Au and ZnO:Pd sensors present response enhancement factors between 2 – 7, compared with the pure ZnO sensor. Among the two sensors with metallic nanoparticles, the ZnO:Pd sensor always shows a higher response compared with the ZnO:Au sensor for similar concentrations, especially at the lower operating temperature of 175$^o$C, in agreement with ZnO sensors decorated with gold and palladium nanoparticles by sputtering [26].

We further explore the properties of the ZnO:Pd sensor by varying the ZnO and Pd deposition parameters. We prepare ZnO thin films, formed by a 3-hour long PLD process, with palladium nanoparticles on the surface, resulting either from a 1-min or a 2-min irradiation of the palladium target. The response of these sensors to acetone is shown in Fig. 4, for various operating temperatures. We note the ZnO:Pd (3h + 2min) sensor (Fig. 4b) allows for the detection of lower acetone concentrations, compared with the ZnO:Pd (3h + 1min) sensor (Fig. 4a). The lowest acetone concentration we detect with the ZnO:Pd (3h + 2min) sensor is 26 ppm at the operating temperature of

200°C, as shown in Fig. 4b. For similar acetone concentrations and operating temperatures, the ZnO:Pd (3h + 2min) sensor presents systematically higher responses compared with the ZnO:Pd (3h + 1min) sensor. Therefore, higher palladium quantity on the ZnO surface, resulting from a longer palladium target irradiation time with all other deposition parameters identical, improves the sensor performance. In order to evaluate the effect of the ZnO film thickness, we compare Fig. 3c with Fig. 4a, which show the response of ZnO films formed by PLD processes of different durations (2 h and 3 h, respectively), with the same amount of palladium on the surface (1 min). For pure ZnO films, the 2-hour PLD process resulted in a thickness of ~185 nm and the 3-hour PLD process in a thickness of ~225 nm. The two sensors show comparable responses for similar acetone concentrations and operating temperatures, indicating the thickness of the ZnO film is not a critical parameter, in contrast with the amount of palladium on the film surface.

From real-time transient response curves, like the one shown in Fig. 2, we can extract the response time of the sensors, which is defined as the time interval between 10% and 90% of the total signal change for each acetone concentration. For all samples, the response time decreases with increasing temperature, reaching the minimum values for the highest operating temperature of 200°C. Also, the addition of metallic nanoparticles on the surface of the films decreases the response time compared with pure ZnO films. For the ZnO:Pd (3h + 2min) sample, which showed the best performance in terms of response magnitude, the response time for the acetone concentrations shown in Fig. 4b at 200°C ranges between 0.4 – 2.4 min.

The decrease of the sensors electric resistance in the presence of acetone is explained by the fact that ZnO is a n-type semiconductor, which in the presence of a reducing gas, such as acetone, gains free electrons which act as majority charge carriers. Molecular atmospheric $O_2$ is initially adsorbed on the surface of ZnO, attracting electrons from ZnO and producing $O_2^-$, $O^-$, and $O^{-2}$ species, depending on the temperature, according to the following reactions [32]:

$$O_2 \text{ (gas)} + e^- \longleftrightarrow O_2^- \text{ (adsorb)} \quad (T \leq 200°C)$$
$$1/2 O_2 \text{ (gas)} + e^- \longleftrightarrow O^- \text{ (adsorb)} \quad (T \geq 200°C)$$
$$1/2 O_2 \text{ (gas)} + 2e^- \longleftrightarrow O^{-2} \text{ (adsorb)} \quad (T \geq 200°C)$$

Due to the transfer of electrons from ZnO to the adsorbed oxygen species, the resistance of ZnO initially increases. Acetone undergoes extreme oxidation by the chemisorbed anionic oxygen species, according to the reactions [33]:

$$CH_3COCH_3 + 4O_2^- \longleftrightarrow 3CO_2 + 3H_2O + 4e^-$$
$$CH_3COCH_3 + 8O^- \longleftrightarrow 3CO_2 + 3H_2O + 8e^-$$
$$CH_3COCH_3 + 8O^{-2} \longleftrightarrow 3CO_2 + 3H_2O + 16e^-$$

The free electrons that result from the above reactions decrease the sensors resistance in the presence of acetone. The gold and palladium nanoparticles form nano-Schottky junctions with ZnO, decreasing the electron concentration in the conduction band of the nanocomposite material compared with pure ZnO [23,34]. The decrease of the conduction band electron concentration renders the nanocomposite sensors more sensitive to small variations of the free electron population and thus more sensitive to the presence of small concentrations of acetone. Additionally, the presence of gold and palladium nanoparticles on the surface of the ZnO films increases the active surface area of the sensor because atmospheric oxygen gets adsorbed on the surface of the nanoparticles and reacts with acetone. Furthermore, metallic nanoparticles act as catalysts, promoting the dissociation of adsorbed oxygen and acetone to highly activated atomic species, which react more efficiently [34]. These mechanisms also explain the decrease of the response time of the sensors in the presence of metallic nanoparticles.

The improved performance of the ZnO:Pd sensors compared with the ZnO:Au sensors is supported by previous studies showing that ZnO:Pd nanostructures attract more chemisorbed oxygen on the surface, compared with ZnO:Au nanostructures. [26]. Additionally, palladium has a slightly higher work function than gold, resulting in a higher Schottky barrier when in contact with ZnO and a more depleted conduction band, which is expected to increase the response of ZnO:Pd gas sensors. Similar results have been obtained for NiO:Pd hydrogen sensors, compared with NiO:Au sensors [35].

## 4. Conclusions

We developed nanocomposite ZnO:Au and ZnO:Pd acetone sensors and compared their performance to each other and to pure ZnO thin-film sensors at relatively low operating temperatures. The nanocomposite sensors were entirely fabricated by PLD, which was employed for the formation of thin ZnO films and partial coverage of their surface with gold or palladium nanoparticles. The ZnO films are polycrystalline, crystallizing mainly at the (002) and (101) orientations of the hexagonal wurtzite phase. The nanocomposite ZnO:Au and ZnO:Pd sensors show a lower detection limit and higher sensitivity than the pure ZnO sensor, especially for lower operating temperatures. The ZnO:Pd sensor performs better than the ZnO:Au sensor. The ZnO:Pd sensor sensitivity increases with the amount of palladium on the surface, while it remains roughly unchanged with the ZnO thickness.

**Acknowledgements**

Financial support of this work by the General Secretariat for Research and Technology, Greece, (project Polynano-Kripis 447963) is gratefully acknowledged.

**References**

1. Z. Wang and C. Wang, Is breath acetone a biomarker of diabetes? A historical review on breath acetone measurements, J. Breath. Res. 7 (2013) 037109.

2. C. Wang and P. Sahay, Breath analysis using laser spectroscopic techniques: breath biomarkers, spectral fingerprints, and detection limits, Sensors 9 (2009) 8230 – 8261.

3. M. Righettoni, A. Tricoli, S.E. Pratsinis, Thermally stable, silica-doped $\varepsilon$-$WO_3$ for sensing of acetone in the human breath, Chem. Mater. 22 (2010) 3152 – 3157.

4. Q.-q. Jia, H.-m. Ji, D.-h. Wang, X. Bai, X.-h. Sun, Z.-g. Jin, Exposed facets induced enhanced acetone selective sensing property of nanostructured tungsten oxide, J. Mater. Chem. A 2 (2014) 13602 – 13611.

5. H. Shan et al., Highly sensitive acetone sensors based on La-doped $\alpha$-$Fe_2O_3$ nanotubes, Sensor. Actuat. B 184 (2013) 243 – 247.

6. X. Sun, H. Ji, X. Li, S. Cai, C. Zheng, Mesoporous $In_2O_3$ with enhanced acetone gas-sensing property, Mater. Lett. 120 (2014) 287 – 291.

7. W.Q. Li et al., Synthesis of hollow $SnO_2$ nanobelts and their application in acetone sensor, Mater. Lett. 132 (2014) 338 – 341.

8. Q. Qi et al., Selective acetone sensor based on dumbbell-like ZnO with rapid response and recovery, Sensor. Actuat. B 134 (2008) 166 – 170.

9. S.H. Yan et al., Preparation of $SnO_2$-ZnO hetero-nanofibers and their application in acetone sensing performance, Mater. Lett. 159 (2015) 447 – 450.

10. X.B. Li, Porous spheres-like ZnO nanostructures as sensitive gas sensors for acetone detection, Mater. Lett. 100 (2013) 119 – 123.

11. D. An, Y. Li, X. Lian, Y. Zou, G. Deng, Synthesis of porous ZnO structures for gas sensor and photocatalytic applications, Colloid. Surface A 447 (2014) 81 – 87.

12. X. Li, Y. Chang, Y. Long, Influence of Sn doping on ZnO sensing properties for ethanol and acetone, Mat. Sci. Eng. C 32 (2012) 817 – 821.

13. C.S. Prajapati and P.P. Sahay, Influence of In doping on the structural, optical and acetone sensing properties of ZnO nanoparticulate thin films, Mat. Sci. Semicond. Proc. 16 (2013) 200 – 210.

14. G.H. Zhang *et al.*, Morphology controlled syntheses of Cr doped ZnO single-crystal nanorods for acetone gas sensor, Mater. Lett. 165 (2016) 83 – 86.

15. F. Tian, Y. Liu, K. Guo, Au nanoparticle modified flower-like ZnO structures with their enhanced properties for gas sensing, Mat. Sci. Semicond. Proc. 21 (2014) 140 – 145.

16. Y. Zeng *et al.*, Growth and selective acetone detection based on ZnO nanorod arrays, Sensor. Actuat. B 143 (2009) 93 – 98.

17. J. Luo *et al.*, The mesoscopic structure of flower-like ZnO nanorods for acetone detection, Mater. Lett. 121 (2014) 137 – 140.

18. S.B. Khan, M. Faisal, M.M. Rahman, A. Jamal, Low-temperature growth of ZnO nanoparticles: Photocatalyst and acetone sensor, Talanta 85 (2011) 943 – 949.

19. H. Bian *et al.*, Improvement of acetone gas sensing performance of ZnO nanoparticles, J. Alloy. Compd. 658 (2016) 629 – 635.

20. S.S. Nath, M. Choudhury, D. Chakdar, G. Gope, R.K. Nath, Acetone sensing property of ZnO quantum dots embedded on PVP, Sensor. Actuat. B 148 (2010) 353 – 357.

21. I. Sta *et al.*, Hydrogen sensing by sol-gel grown NiO and NiO:Li thin films, J. Alloy. Compd. 626 (2015) 87 – 92.

22. X.-j. Wang, W. Wang, Y.-L. Liu, Enhanced acetone sensing performance of Au nanoparticles functionalized flower-like ZnO, Sensor. Actuat. B 168 (2012) 39 – 45.

23. Y. Lin *et al.*, Highly stabilized and rapid sensing acetone sensor based on Au nanoparticle-decorated flower-like ZnO microstructures, J. Alloy. Compd. 650 (2015) 37 – 44.

24. M. Kandyla, C. Chatzimanolis-Moustakas, E.P. Koumoulos, C. Charitidis, M. Kompitsas, Nanocomposite NiO:Au hydrogen sensors with high sensitivity and low operating temperature, Mater. Res. Bull. 49 (2014) 552 – 559.

25. I. Sta *et al.*, Surface functionalization of sol-gel grown NiO thin films with palladium nanoparticles for hydrogen sensing, Int. J. Hydrogen Energ. 41 (2016) 3291 – 3298.

26. Y.-C. Liang, W.-K. Liao, X.-S. Deng, Synthesis and substantially enhanced gas sensing sensitivity of homogeneously nanoscale Pd- and Au-particle decorated ZnO nanostructures, J. Alloy. Compd. 599 (2014) 87 – 92.

27. R. Khandelwal *et al.*, Effects of deposition temperature on the structural and morphological properties of thin ZnO films fabricated by pulsed laser deposition, Optics & Laser Technology 40 (2008) 247 – 251.

28. A. Klini, A. Manousaki, D. Anglos, C. Fotakis, Growth of ZnO thin films by ultraviolet pulsed-laser ablation: Study of plume dynamics, Journal of Applied Physics 98 (2005) 123301.

29. I. Fasaki, M. Kandyla, M. Kompitsas, Properties of pulsed laser deposited nanocomposite NiO:Au thin films for gas sensing applications, Appl. Phys. A 107 (2012) 899 – 904.

30. I. Fasaki, M. Kandyla, M.G. Tsoutsouva, M. Kompitsas, Optimized hydrogen sensing properties of nanocomposite NiO:Au thin films grown by dual Pulsed Laser Deposition, Sensor. Actuat. B 176 (2013) 103 – 109.

31. E. Gyorgy, J. Santiso, A. Figueras, A. Giannoudakos, M. Kompitsas, I.N. Mihailescu, Morphology evolution and local electric properties of Au nanoparticles on ZnO thin films, Journal of Applied Physics 98 (2005) 84302.

32. N. Barsan and U. Weimar, Conduction model of metal oxide gas sensors, J. Electroceram. 7 (2001) 143 – 167.

33. M. El-Maazawi, A.N. Finken, A.B. Nair, V.H. Grassian, Adsorption and photocatalytic oxidation of acetone on $TiO_2$: an in situ transmission FT-IR study, J. Catal. 191 (2000) 138 – 146.

34. A. Kolmakov, D.O. Klenov, Y. Lilach, S. Stemmer, M. Moskovits, Enhanced gas sensing by individual $SnO_2$ nanowires and nanobelts functionalized with Pd catalyst particles, Nano Lett. 5 (2005) 667 – 673.

35. M. Kandyla, C. Chatzimanolis-Moustakas, M. Guziewicz, M. Kompitsas, Nanocomposite NiO:Pd hydrogen sensors with sub-ppm detection limit and low operating temperature, Mater. Lett. 119 (2014) 51 – 55.

**Figure Captions**

**Figure 1:** X-ray diffractogram of a ZnO thin film, deposited by 2-hour PLD.

**Figure 2:** Transient response of a ZnO:Pd sensor to acetone, for an operating temperature of 185°C. The ZnO film is formed by 3-hour PLD and the palladium nanoparticles by 2-min PLD.

**Figure 3:** Sensor response as a function of acetone concentration for (a) a pure ZnO film, formed by 2-hour PLD, (b) a similar ZnO film with gold nanoparticles on the surface, formed by 1-min PLD, and (c) a similar ZnO film with palladium nanoparticles on the surface, formed by 1-min PLD. The sensors operate at 175°C and 200°C.

**Figure 4:** Sensor response as a function of acetone concentration for a ZnO film, formed by 3-hour PLD, with palladium nanoparticles on the surface, formed by (a) 1-min and (b) 2-min PLD. The sensors operate at the indicated temperatures.

# Figures

## Figure 1

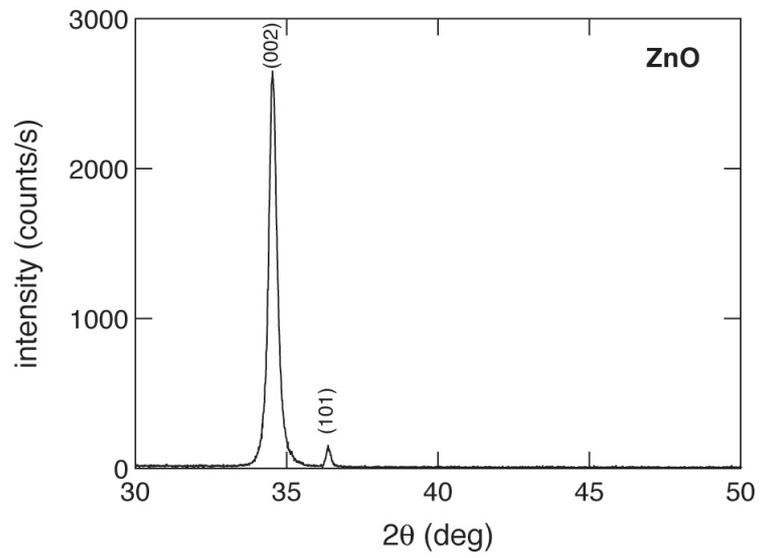

## Figure 2

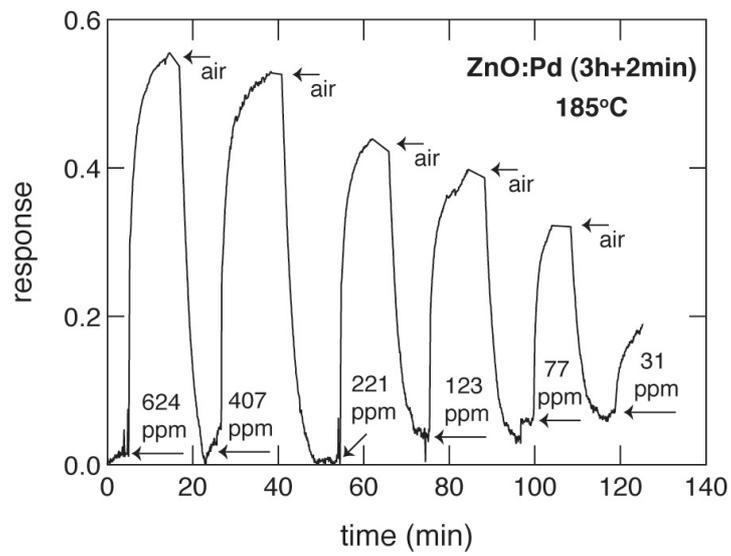

**Figure 3**

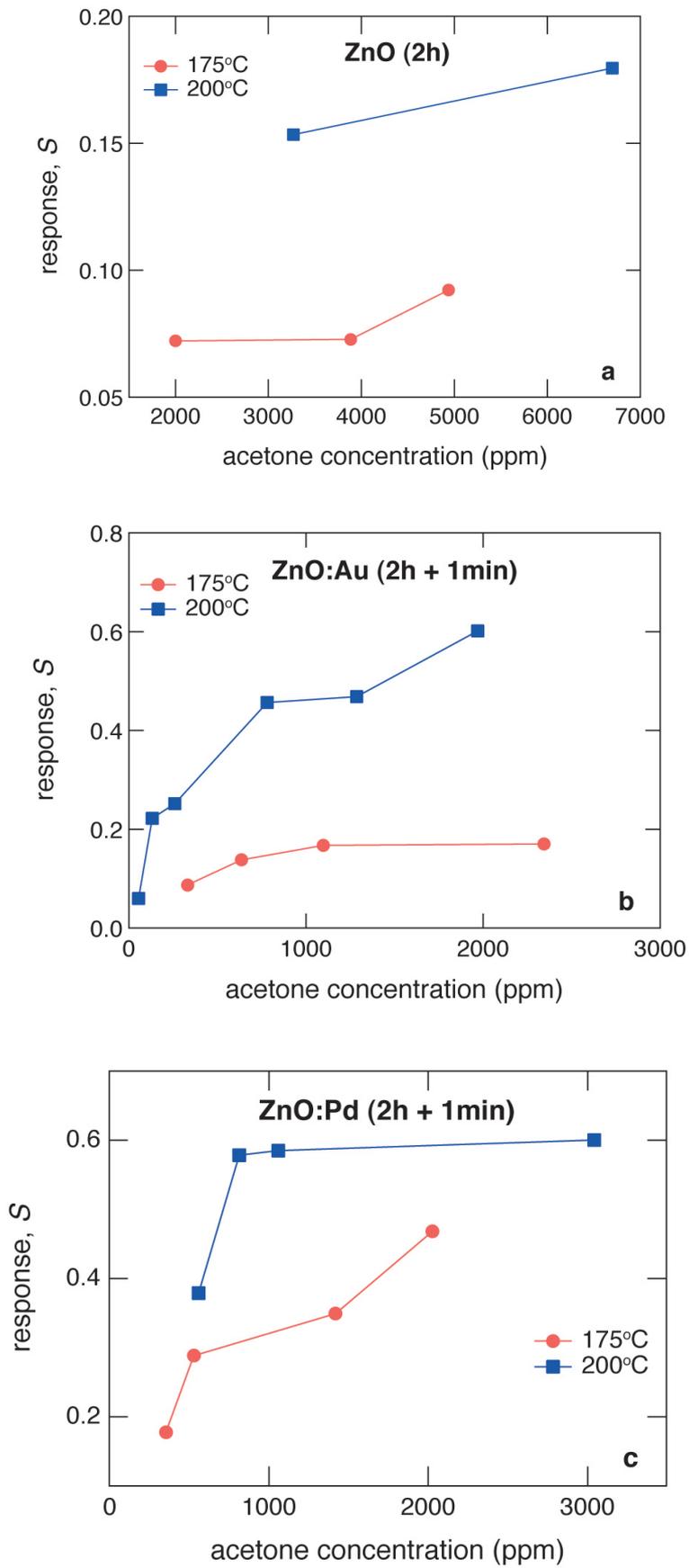

**Figure 4**

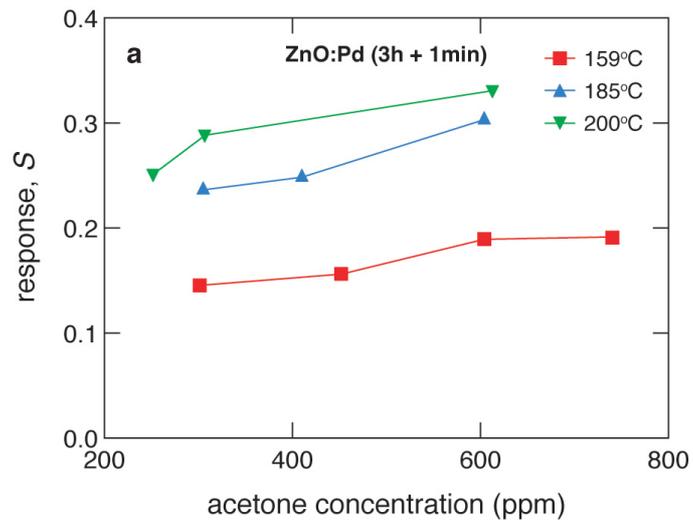

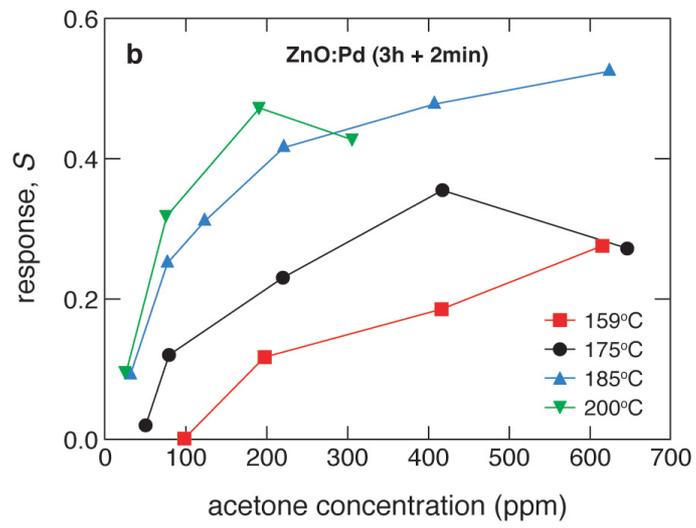

# Supplementary Material

# Pulsed laser deposition of ZnO thin films decorated with Au and Pd nanoparticles with enhanced acetone sensing performance

*M. Alexiadou, M. Kandyla, G. Mousdis, and M. Kompitsas*

Energy-dispersive x-ray spectroscopy measurements (EDS) provide information about the composition of the samples, especially after Au or Pd deposition. Figure S1 shows SEM images of the ZnO:Au (2h+1min) and ZnO:Pd (3h+2min) samples and the corresponding areas on which EDS spectra were obtained. Figure S2 shows the EDS spectrum of the ZnO:Au (2h+1min) sample and Fig. S3 shows the EDS spectrum of the ZnO:Pd (3h+2min) sample. The composition of the samples is shown in Tables S1 and S2.

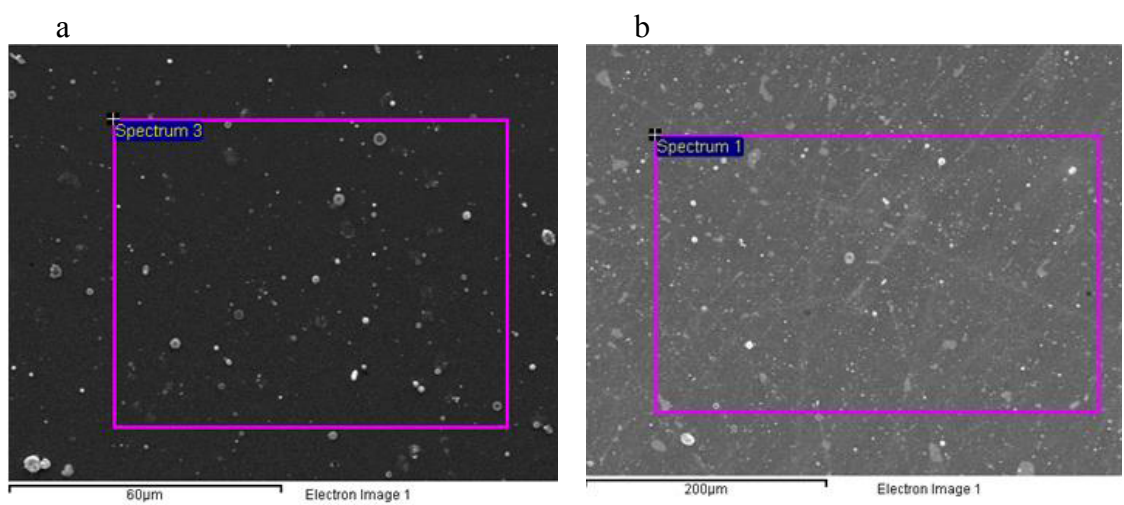

**Figure S1:** SEM images of (a) ZnO:Au (2h + 1min) and (b) ZnO:Pd (3h + 2min) samples. EDS spectra were obtained from the indicated areas.

---

. Corresponding author: kandyla@eie.gr

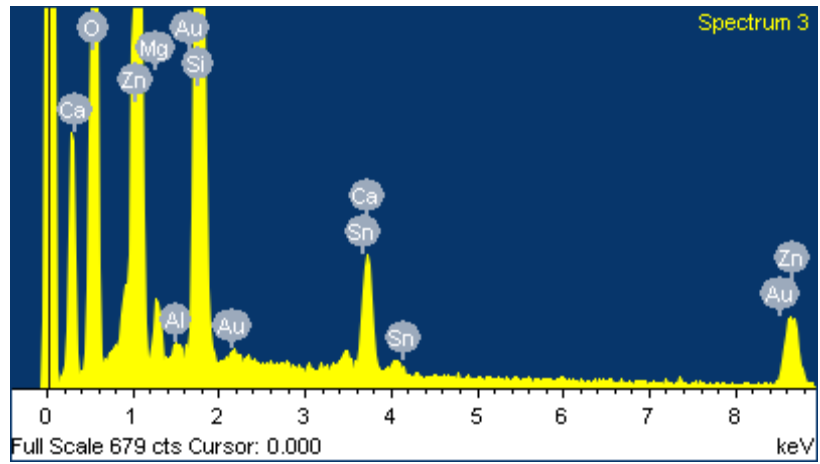

**Figure S2:** EDS spectrum of the ZnO:Au (2h+1min) sample.

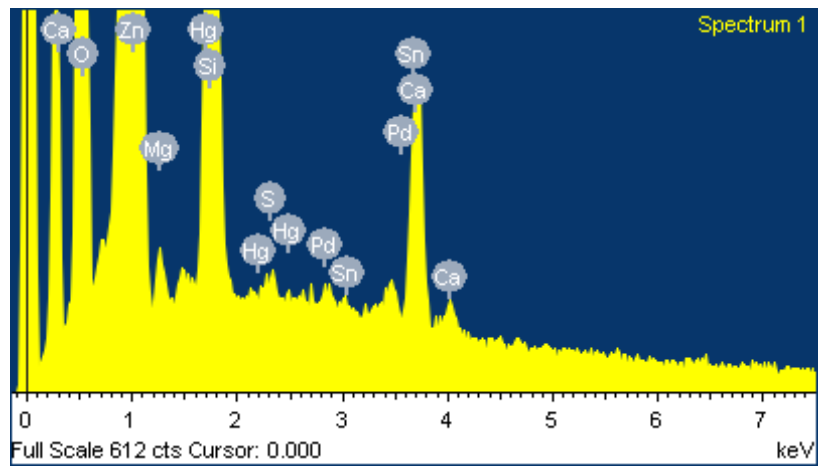

**Figure S3:** EDS spectrum of the ZnO:Pd (3h+2min) sample.

| Element | Weight% | Atomic% |
|---------|---------|---------|
| O K     | 19.94   | 58.19   |
| Mg K    | 0.96    | 1.85    |
| Al K    | 0.14    | 0.24    |
| Si K    | 13.5    | 22.45   |
| Ca K    | 2.19    | 2.55    |
| Zn L    | 19.95   | 14.25   |
| Sn L    | 0.93    | 0.37    |
| Au M    | 0.42    | 0.1     |

**Table S1:** Composition of ZnO:Au (2h+1min).

| Element | Weight% | Atomic% |
|---|---|---|
| O K | 23.20 | 55.10 |
| Mg K | 0.59 | 0.92 |
| Si K | 11.19 | 15.15 |
| S K | 0.15 | 0.17 |
| Ca K | 2.13 | 2.02 |
| Zn L | 45.13 | 26.23 |
| Pd L | 0.40 | 0.14 |
| Sn L | 0.79 | 0.25 |
| Hg M | 0.03 | 0.01 |

**Table S2:** Composition of ZnO:Pd (3h+2min).